\documentclass[
    ,final            
  ]
  {aipproc}

\layoutstyle{6x9}

\def\Msun{$M_\odot$}

\def\simgt{\lower.5ex\hbox{$\; \buildrel > \over \sim \;$}}
\def\simlt{\lower.5ex\hbox{$\; \buildrel < \over \sim \;$}}
\def\Porb{P$_{\rm orb}$}
\def\Pbif{P$_{\rm bif}$}

\begin{document}

\title{Were most Low Mass X ray Binaries born in Globular Clusters?}

\classification{97.60.Gb, 97.80.Jp, 98.20.Gm 
                }
\keywords      {stars: millisecond pulsars, low mass X-ray binaries, globular clusters}

\author{Francesca D'Antona}{
  address={INAF-Osservatorio di Roma, via di Frascati 33, I-00040 Monteporzio (Italy)}
}

\author{Anamaria Teodorescu}{
  address={Institute for Astronomy, University of Hawaii, 2680 Woodlawn Drive, Honolulu, 
HI 96822}
}

\author{Paolo Ventura}{
  address={INAF-Osservatorio di Roma, via di Frascati 33, I-00040 Monteporzio (Italy)}
  }

\begin{abstract}
We summarize the status of art of the secular evolution of low mass X--ray binaries (LMXBs)
and take a close look at the orbital period distribution of LMXBs and of binary
millisecond pulsars (MSP), in the hypothesis that this latter results from the LMXB
evolution. The deficiency of systems below the period gap, 
which in cataclysmic binaries occurs between $\sim 2$ and
3~hr, points to a very different secular evolution of LMXBs with respect to their 
counterparts containing a white dwarf compact object. The presence of several
ultrashort period LMXBs (some of which are also X--ray millisecond pulsars), 
the important fraction of binary MSPs at periods between 0.1 and 1 day, the periods 
(26 and 32hr) of two ``interacting'' MSPs in Globular Clusters are other pieces of the
puzzle in the period distribution. We consider the possible explanations for
these peculiarities, and point out that Grindlay's old proposal that all (most of) 
LMXBs in the field were originally born in globular clusters must 
be carefully reconsidered.
\end{abstract}

\maketitle

\section{Introduction}

We re--examine the secular evolution of low mass
X ray binaries (LMXBs) and its possible remnant population, the millisecond pulsars
(MSP) both in the field and in Globular Clusters (GCs). In particular 
we pay attention to the possible effects 
of a ``radio--ejection" phase, initiated when the mass transfer temporarily stops 
during the secular evolution. 

The fraction of binary MSPs and LMXBs in GCs is much larger 
with respect to the fraction in the galactic field. This is  
regarded as a clear indication that binaries containing neutron stars (NS) in GCs are
generally not primordial, but are a result of stellar encounters due to
the high stellar densities in the GC cores. On the other hand, it is still not
clear how the LMXBs are formed in the galactic field. In fact, the supernova
explosion in a binary in which the companion is a low mass star will generally 
destroy the binary. Possible paths have been suggested to explain the survival of
LMXBs in the field: 
\begin{enumerate}
\item accretion induced collapse of a white dwarf primary into a neutron stars
\cite{flannery1975,canal1980}; 
\item supernova ``kicks" due to asymmetric neutrino energy deposition 
during the supernova event \cite{kalogera1998}; 
\item a large fraction of today's LMXBs could have started their life as intermediate 
mass X-ray binaries (IMXBs) stars, with donors initially as massive as 2-3\Msun
\cite{podsi2002}; 
\item LMXBs were formed 
by capture in the dense environment of Globular Clusters which, later on, were
destroyed, and the binary remains isolated in the galactic field or bulge \cite{grindlay1984}.
\end{enumerate}
This latter hypothesis, originally proposed by Grindlay in 1984, 
was recently re--appraised by Podsiadlowski et al.
\cite{podsi2002} and is also proposed in this talk, based
on some evidence coming from the period distribution of LMXBs.
Further observations in favour of this hypothesis, for
early type galaxies, comes from the finding that the
ratio of the global LMXB X--ray luminosity to galactic optical luminosity in
elliptical galaxies is strongly correlated with the specific globular cluster frequency 
in ellipticals \cite{white32002, kimfabbiano2004}. A general discussion of the origin of
LMXBs in external galaxies is given in \cite{fabbiano2006}. 

In the following we examine both the evolution above and below the
``bifurcation period" \Pbif\ \cite{tutukov1985, pylyser1988}
and discuss what we can infer from the comparison of the
resulting orbital period distributions with the observed ones.

\section{Evolution starting at P$>$\Pbif}

The ``standard" secular evolution of LMXBs as progenitors of binary MSPs finds 
a reasonable application to the evolution of binaries above the so called 
``bifurcation period" \Pbif\ \cite{tutukov1985, pylyser1988}, in which
the donor stars begins the mass transfer phase after it has finished the
phase of core hydrogen burning, and the system ends up as a low mass white
dwarf (the remnant helium core of the donor) in a relatively long
or very long orbit with a radio MSP (e.g. \cite{rappaport1995}).
Some of these systems may also be the remnant of the evolution 
of intermediate mass donors, with similar resulting orbital periods, see, e.g., 
\cite{podsi2002}. Recently, \cite{dantona2006} we have shown that the secular evolution 
at P$>$\Pbif\ may need to take into account the detailed stellar evolution
of the giant donor, to explain the orbital period gap of binary MSPs between
$\sim 20$\ and $\sim 60$ days. During the evolution along the RGB, the 
hydrogen burning shell encounters the hydrogen chemical discontinuity 
left by the maximum deepening of convection \cite{thomas}. 
The thermal readjustment of the shell causes a luminosity and radius drop, which
produces a well known ``bump" in the luminosity function of the red giant branch
in globular clusters \cite{ferraro1999, zoccali}. In semi detached binaries, at the ``bump",
the mass transfer has a temporary stop, due to the sudden 
decrease of the radius. When the radius increases and the mass transfer starts again, 
we consider possible that a phase of ``radio--ejection" begins 
(\cite{burderi2001, burderi2002}), in which mass accretion on the neutron
stars is not allowed due to the pressure exerted by the radio pulsar. In this case,
the matter is lost from the system at the inner lagrangian point, carrying away
angular momentum and altering the period evolution. This will occur especially for magnetic 
moments of the NS in a range ($2 \times \simlt \mu \simlt 4) \times 10^{26}$Gcm$^3$,
which is the most populated range for binary MSPs. If this is the explanation,
we predict that the magnetic moments of binary MSPs which
can be eventually found in the period gap at 20--60 days should be 
preferentially either smaller or larger than the range quoted above.

\begin{figure}
  \includegraphics[height=.4\textheight]{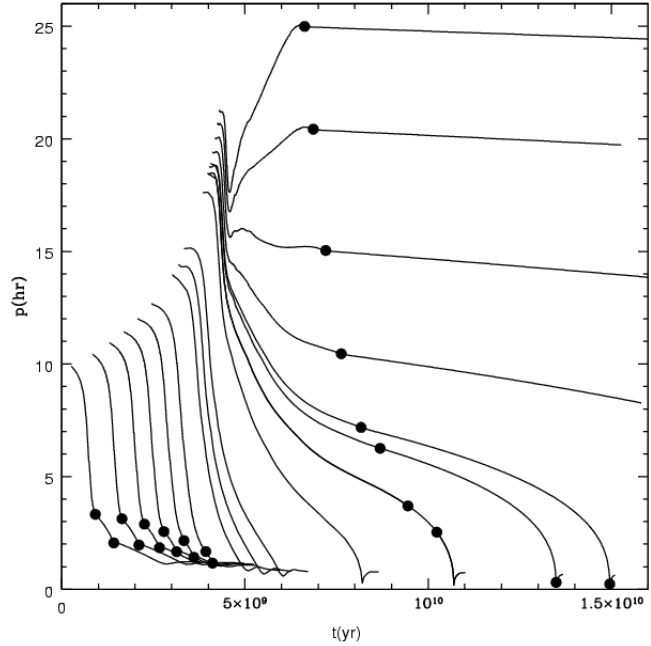}
  \caption{The evolutions in the plane \Porb\ vs. age for donors of 1.2\Msun, Z=0.01
  starting mass loss from different initial \Porb, and thus different degrees of hydrogen
  consumption in the stellar core. The dots indicate phases at which mass transfer stops. 
  If present, the following dot along the same evolution indicates where mass transfer resumes.
  The scarcely evolved sequences (those starting from the shorter \Porb) all show a more
  or less wide period gap at 2--3hr. The more evolved sequences do not show a period gap and
  reach very short \Porb: these can explain the ultra--short period LMXBs at \Porb$\sim$40m. 
  Two among the
  even more evolved sequences detach at \Porb$\simlt$6hr, so that they can resume mass
  transfer within the Hubble time, and go down to the shortest \Porb systems (11 and 20m).
  The others detach at the intermediate periods below $\sim$1 day 
  where many binary MSPs are found.
  }
  \label{f1}
\end{figure}

\section{Evolution starting at P$<$\Pbif}

If the secular evolution 
of LMXBs below \Pbif\ is similar to the evolution of cataclysmic binaries (CBs), 
we should expect many systems at P$\simlt$2~hr, and  
a minimum orbital period similar to that of CBs, namely $\sim$80 minutes.
On the contrary, there are very few of these systems, and there are several ``ultrashort"
binaries. In particular, three LMXBs in the field (which are also X--ray MSPs) and
one in a Globular Cluster (XB~1832-330 in NGC~6652 \cite{deutsch2000}) 
are concentrated in the range 44$\simlt$ \Porb $\simlt 40$ minutes 
\cite{markwardt2002, galloway2002,
markwardt2003ATel27}. In addition there are two other ultrashort
systems in Globular Clusters (XB~1850-087 in NGC 6712 at \Porb=20.4m and
XB~1820-303 in NGC 6624 at \Porb=11.4min). 
While for the systems in GCs we may think that they
were formed by capture of a white dwarf by the neutron star, the systems in the field 
should have reached this phase by secular evolution. Models have been built up
\cite{nelsonrappaport2003, podsi2002}, that imply that the donors
began Roche lobe overflow at periods {\it just a bit} below \Pbif,
so that the donor evolved to become a degenerate dwarf
predominantly composed of helium, but having a residual, very small ($<$10\%) 
hydrogen abundance.
Until hydrogen is present in the core of the donor star, in fact, the evolution 
proceeds towards short \Porb (convergent systems). 
The mass radius relation when these objects are close to degeneracy, if the hydrogen content
left is very small, is intermediate between that of hydrogen dominated brown dwarfs and that 
of helium white dwarfs, so that smaller radii and 
shorter \Porb\ will be reached before radius and period begin increasing again.
One problem of this scenario is that {\it there is a very small interval of
initial \Porb} which allows this very peculiar evolution: in most cases, either
a helium core is already formed before the mass transfer starts, and the system
evolves towards long \Porb, or there is enough hydrogen that the
system is convergent, but the minimum period is similar to that of CBs, and can not
reach the ultrashort domain (see also \cite{sluys2005}).
This is shown in Figure \ref{f1} \cite{anamaria}, 
where the evolutions for a donor of 1.2\Msun, 
metallicity Z=0.01 in mass fraction, are shown for increasing starting periods for
the mass loss phase.
\begin{figure}
  \includegraphics[height=.4\textheight]{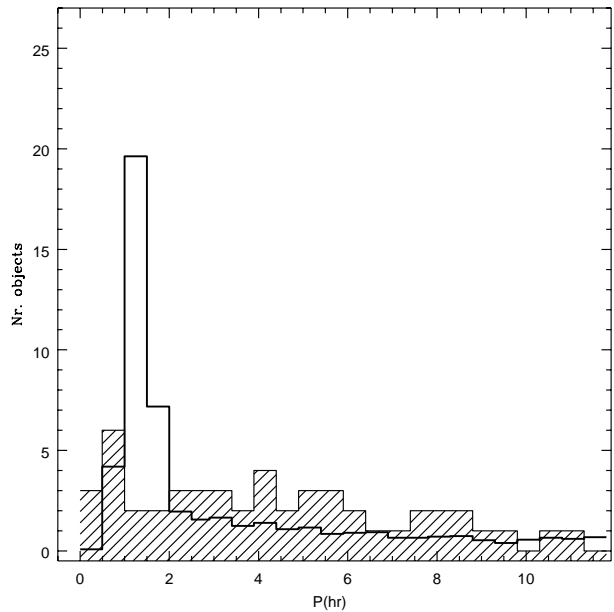}
  \caption{Number vs. period distribution expected for LMXBs, by applying a population
  synthesis analysis based on a library of evolutions similar to those presented
  in Figure \ref{f1}. The analysis is from \cite{anamaria}. }
  \label{f2}
\end{figure}

Podsiadlowski et al. \cite{podsi2002} notice that, for a 1\Msun\ secondary, 
the initial period range 13--18~hr leads to the formation of ultracompact systems. 
Since systems that start mass transfer in this period range are naturally 
produced as a result of tidal capture, this may explain the large
fraction of ultracompact LMXBs observed in globular clusters. This however does not
apply to LMXBs in the field. In her phD thesis, A. Teodorescu \cite{anamaria} built
the number vs. period distribution expected for LMXBs for convergent systems
under several hypotheses, and compared it with the available observed 
period distribution. An expected result was that the range \Porb$<2$hr 
is very populated (Figure \ref{f2})
and the distribution is inconsistent with observations, {\it unless we can suppress the 
secular evolution of all the systems below the ``period gap"} which occurs
at about the same location as in CBs, as shown in Figure \ref{f1}. 
We can explore several different possibilities to do this: 
\begin{figure}
\includegraphics[height=.4\textheight]{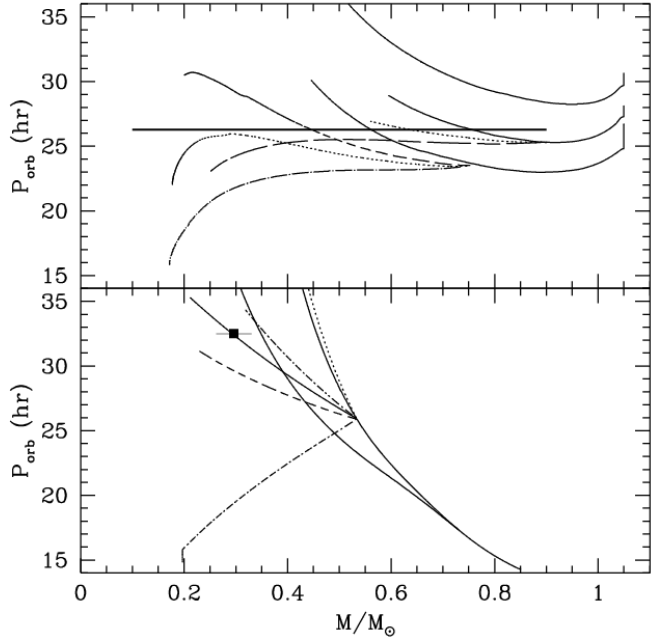}
  \caption{In the bottom figure we show
  the possible evolutions for the companion of PSR J1740-5340.
  All tracks start close to the turnoff of a 
  M=0.85M$_\odot$, when hydrogen is almost fully depleted in the central 
  regions. The continuous line evolving to the longest periods is the 
  conservative evolution considered in \cite{burderi2002}. The curve passing 
  through the square indicating the observed mass of the donor shows the 
  evolution obtained by assuming that, starting at a period of $\sim 26$hr, the 
  mass is lost from the system with the specific angular momentum of the 
  donor star. The other curves starting from the same P=26hr represent evolutions
  with different amounts of loss of specific angular momentum.
  The top figure shows similar evolutions starting from a donor of 1.1\Msun, aimed
  to interpret the evolutionary status of the donor of PSR J1748-2446ad in the
  GC Terzan 5. A solar chemistry is assumed here, close to the
  metallicity of this cluster. The donor mass is not 
  observationally constrained here, but it looks evident again that
  the evolution of the donor must have started close to the bifurcation period.}
  \label{f3}
\end{figure}
\begin{enumerate}
\item It is possible that the lack of \Porb's $<$2~hr is again 
an effect due to radio--ejection: after the period
gap is traversed by the detached system, when the mass transfer resumes, accretion
is prevented by the pulsar pressure, the matter escapes from the system with high
specific angular momentum, and the evolution is accelerated. This is possibly
occurring in the system containing the radio pulsar PSR J0024.7204W 
in the GC 47 Tuc, at \Porb=3.2~hr. This system exhibits
X-ray variability which can be explained by the
presence of a relativistic shock within the binary that is regularly
eclipsed by the secondary star \cite{bogdanov}. 
The shock can then be produced by the interaction
of the pulsar wind with a stream of gas from the companion
issuing through the inner Lagrange point (L1), a typical case of what is intended
by radio--ejection \cite{burderi2001}.
This mechanism could affect all the systems which enter a period gap. Notice that
only systems which end up at ultra--short periods {\it do not} detach during the
secular evolution: is this the reason why they are so prominent in the number vs. period
distribution? In the framework of this explanation, both the lack of systems 
at \Porb$<2$~hr and the presence of ultrashort periods would be due to this effect.
\item ``Evaporation" of the donor, due to the the pulsar wind impinging on,
and evaporating material off of the surface of the companion (e,g, \cite{rudermantavani})
is another possible mechanism. The difference with respect to the previous case
is that evaporation is effective also during the phase in which the system is detached.
\item It is possible that the secular evolution practically never begins when the donor is
not significantly evolved. This can be true only if binaries are mostly 
formed by tidal capture, in which the NS captures a main sequence object only at
separations 1.5R$_* \simlt a \simlt 3 R_*$ \cite{fabian1975}, as it may happen in 
GCs, but we are examining the \Porb\ distribution of {\it all} the LMXBs in the 
Galaxy. We should then {\it re--appraise Grindlay's suggestion that {\it most} of the 
field LMXBs were formed in GCs, which were later on destroyed} by tidal interactions 
with the galactic bulge.
\end{enumerate}
At the time of writing up this talk, we still have the opinion that the provocative 
issue that most LMXBs were actually born in GCs must be considered again in more 
detail.

\section{Evolution starting at P very close to \Pbif}

There are other cases which show us that some evolutions
start very close to \Pbif: the famous interacting MSP binary
in NGC 6397, PSR~J1740-5340 is such a case: at \Porb=35.5~hr, it seems to be in a
radio--ejection phase \cite{burderi2002}. Surely the companion has not 
been captured recently by stellar encounters: it is an evolved subgiant, as
predicted by the secular evolution models \cite{burderi2002}, 
and as testified by the CN cycled chemistry of the donor envelope \cite{sabbi2003}, 
and predicted by Ergma \& Sarna \cite{ergma-sarna2003}. 
We suggest that also PSR J1748-2446ad, the fastest
spinning MSP discovered in the Globular Cluster Terzan 5 \cite{hessels2006} 
is in a radio--ejection phase \cite{burderi2006}, but the
lack of information on the donor preclude a very secure interpretation. 
At \Porb=26.26~hr, again, the donor should be in an early subgiant stage, and
have evolved very close to \Pbif. 
In Figure \ref{f3} we show the possible period evolution for the companion of
PSR~J1740-5340, which must be indeed very slightly evolved off the main sequence 
in order to fit the orbital period and the donor mass very well determined from 
observations, and the possible evolutionary paths for the donor of PSR J1748-2446ad,
which must be even less evolved at the beginning of mass transfer, both because the
final period is shorter (26.6~hr) and because Terzan 5 has a larger metallicity than 
NGC 6397, so that the initial stellar radii are larger for a mass evolving in a time
span of about 10$^{10}$yr. Unfortunately, nothing is known about the optical
counterpart of this system, so that the evolution can not be determined better.

Finally, we have to remember that {\it the whole period distribution 
of binary MSPs in GCs} is consistent with a very high
probability of beginning of mass exchange close to \Pbif: in fact,
there is a large group having \Porb\ from 0.1 to 1 days, 
a range, e.g., not covered at 
all by the ``standard" evolutions by \cite{podsi2002}, but which results easily from
the range of initial periods which are between those leading to ultrashort binaries
and those above bifurcation (see Figure \ref{f1} from \cite{anamaria}). 
Finally, there are several binary MSPs in GCs
for which the white dwarf mass is very low (0.18--0.20\Msun), close to the minimum mass
which can be formed by binary evolution \cite{burderi2002}, indicating again
evolution starting at a period slightly larger than \Pbif.


%


\bibliographystyle{aipproc}   

\end{document}